\documentclass[%
 reprint,
 amsmath,amssymb,
 aps,
prb, citeautoscript,
]{revtex4-1}

\usepackage{graphicx}
\usepackage{dcolumn}
\usepackage{bm}
\usepackage{color, soul}
\usepackage{natbib}
\usepackage[colorlinks, linkcolor=blue, citecolor=blue, urlcolor=blue]{hyper ref}
\makeatletter
\newcommand*{\rom}[1]{\expandafter\@slowromancap\romannumeral #1@}
\makeatother

\begin{document}

\preprint{APS/123-QED}

\title{Density Functional Theory Study of Surface Stability and Phase Diagram of Orthorhombic CsPbI$_3$}

\author{Kejia Li}
\author{Mengen Wang}
\email{mengenwang@binghamton.edu}
\affiliation{%
Department of Electrical and Computer Engineering and Materials Science and Engineering Program, SUNY Binghamton, NY 13902, USA
}%

\begin{abstract}
CsPbI$_3$ has been recognized as a promising candidate for optoelectronic device applications.
To further improve the efficiency of the devices, it is imperative to better understand the surface properties of CsPbI$_3$, which affect charge carrier transport and defect formation properties.
In this study, we perform density functional theory calculations to explore the stability of the (001), (110), and (100) surfaces of orthorhombic CsPbI$_3$, considering different stoichiometries and surface reconstructions.
Our results show that, under the chemical potentials confined by the thermodynamically stable region of bulk CsPbI$_3$, the CsI-terminated surfaces of (001) and (110) and the stoichiometric surface of (100) are stable.
Among these three surfaces, the CsI-terminated (110) surface has the lowest surface energy and no mid-gap states, which benefits the transport properties of the material.

\end{abstract}

\maketitle

\section{\label{intro}Introduction}

Perovskite solar cells (PSCs) are known for their simple synthesis methods, chemically tunable compositions, and design flexibility, which contribute to the potential for high photovoltaic efficiency.\cite{Sharif_2023_pscreview,Snaith_2013_pscreportnewera,Kim_2014_organpscreport} 
Pb-based perovskites are widely investigated for PSCs due to their properties including long carrier lifetimes, convenient bandgap engineering by alloying, high optical absorption coefficients, and low-temperature solution synthesis methods.\cite{Kanij_2022_Pb/Snbasedpscapply,Heo_2016_mapbi/brsolarcellapply}
CsPbI$_3$ with a band gap of 1.7 eV is recognized for its excellent potential for high-efficiency solar cells, which has led to power conversion efficiencies (PCE).\cite{Yu_2021_cspbi320perPCIE, Li_2018_expbandgapincspbi3, Kebede_2023_cspbi3sysn_delta/gammaphasetrans} 

To further improve the performance of the material and device, it is important to understand the surface properties of CsPbI$_3$, which play important roles in material stability and defect engineering strategies.
Point defects on surfaces may trap charge carriers, resulting in efficiency losses.\cite{Huang_2018_cspbi3pointdefect_DFTformEdiffphases, Wang_2019_perovskite_surfacedefect_PCE21-22.6}
Passivation methods of perovskite surfaces have been extensively studied to improve the PCE of the solar cells.\cite{Jia_2021_cspbi3nanodot_psc15-16}
Surface structures and defect dynamics of Pb-based perovskite have been investigated by scanning tunneling microscopy (STM).\cite{Zhang_2022_stmreview_metalhalidepsc, QiSTMmapbbr, MAPbI3stm, zhongSTM} 
The MABr or MAI-terminated surfaces are found to cover the majority of the (001) surface of orthorhombic MAPbBr$_3$\cite{QiSTMmapbbr} and MAPbI$_3$\cite{MAPbI3stm}, and (010) of mixed-halide perovskites\cite{hieulle2019stm_mixed}. 
On MA-halogen-terminated surfaces, both zigzag and dimer patterns are observed due to different orientations of the MA cations.\cite{QiSTMmapbbr,MAPbI3stm} 

Density Functional Theory (DFT) calculations have been applied to investigate the surface structures of Pb-based perovskites, including surface stability, defect formation, and surface passivation.\cite{Xue_2022_perovskitedefectcal_APbX3, Haruyama_2014_mapbi3vacant_dftcal, Yang_2021_cspbbr3_nonpolarcomp_DFT, Han_2022_passivationcspbi3_FtoI}
DFT calculations of the flat surfaces of CsPbI$_3$ show that the CsI-terminated surface is more stable than the PbI$_2$-terminated one under equilibrium growth conditions for the cubic and orthorhombic phases.\cite{long_2019_cspbi3001sur_csipbicompare_vacancys, Seidu_2021_cspbi3001surface_diffdefects}
Experimentally MABr vacancies are observed on the MAPbBr$_3$ (001) surfaces, indicating the material surface can be stable under various stoichiometries.\cite{liu2017mabrbr3_stm, stecker2019mabrbr3_stm}
DFT study on the (110), (001), (100), and (101) surfaces of MAPbI$_3$ shows that compared to the flat PbI$_2$ termination, vacant termination is more stable on all of the surfaces under the thermodynamic equilibrium conditions of the bulk phase.\cite{Haruyama_2014_mapbi3vacant_dftcal}
Our previous work also found that CsSnI$_3$ surfaces with cation vacancies are stable under the thermodynamic equilibrium condition of bulk CsSnI$_3$.\cite{li2024cssni3_surface_dft}
Surface reconstructions with various stoichiometries are commonly observed on semiconductor surfaces, which play important roles in the growth and application of the materials.\cite{timmermann2020iro,wang2020role,dreyer2014absolute,wang2021adsorption,noordhoek2024accelerating,wang2023surface,wang2021incorporation}

To better understand the surface properties of CsPbI$_3$ and explore their stabilities under different thermodynamic conditions, we perform DFT calculations to investigate the (001), (110), and (100) surfaces of orthorhombic CsPbI$_3$ and explore the nature of reconstructions for different stoichiometries.
Supercells are constructed to consider surface reconstructions.
46 different surface structures are built, covering various stoichiometries including Cs-I-rich, Pb-I-rich, Pb-rich, Cs-rich, I-rich, and stoichiometric surfaces. 
The Grand potential method is used to compare the stability of these surfaces under different chemical potentials. 
We also release the code that can read the formation energies of the systems, determine the stable surface stoichiometries under different combinations of chemical potentials, and generate the surface phase diagrams.
From the phase diagram of all surfaces, we determine the most stable surface under the thermodynamic equilibrium condition of the bulk phase is the CsI-terminated flat surface on (001) and (110), and the stoichiometric surface on (100).
Additionally, we find that the stoichiometric surfaces of (001) and (110) also have relatively low surface energies, which are generated by creating CsI vacancies on CsI-terminated flat surfaces.
The electronic properties of the stable structures under thermodynamic equilibrium conditions are also analyzed.

\section{\label{sec:level1}Computational Methods}

\subsection{\label{sec_dft} DFT calculations}

DFT calculations are performed using the projector augmented-wave (PAW) method as implemented in Vienna Ab initio Simulation Package (VASP)\cite{Kresse_1993_VASPinbeginning, Kresse_1994_VASPsemi, Kresse_1996_VASPPAW, Kresse_V1996_VASPPAWsemi}. 
The Perdew–Burke–Ernzerhof (PBE) functional and a 400 eV plane-wave cutoff are used.\cite{PE_1994_PAWmethod,Perdew_1996_GGAmethod} 
The Brillouin zone of the orthorhombic unit cell is sampled using a 2 $\times$ 2 $\times$ 2 $\Gamma$-centered k-mesh. 
For structural relaxations and surface energy calculations, the atoms are relaxed until the forces are below 0.01 eV/\text{\AA}.
The calculated lattice constants of CsPbI$_3$ are \(a = 8.69 \, \text{\AA}\), \(b = 9.12 \, \text{\AA}\), and \(c = 12.64 \, \text{\AA}\), which are in reasonable agreement with the experimental values \(a = 8.86 \, \text{\AA}\), \(b = 8.58 \, \text{\AA}\), and \(c = 12.47 \, \text{\AA}\).\cite{Sutton_2018_cspbi3cubic/orthphase_transition_expdata}

\begin{figure}
\includegraphics[width=85mm]{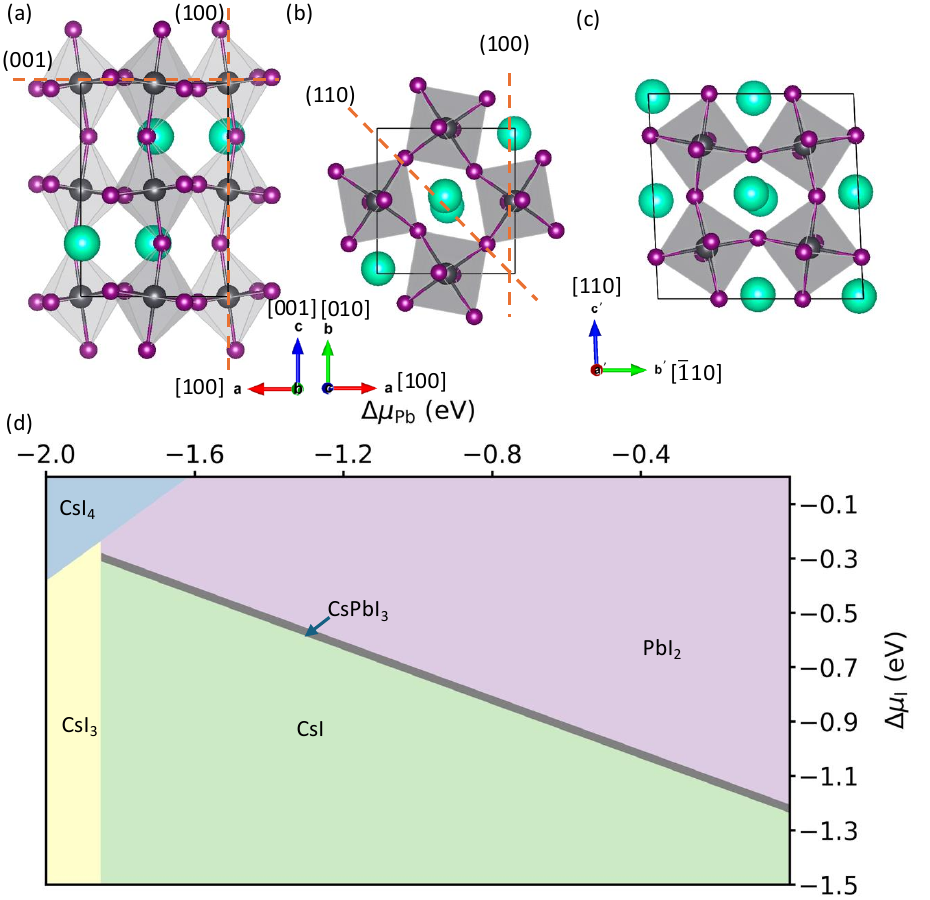}
\caption{\label{fig:phasediagram} (a) Side view and (b) top view of the structure and surface cut of the orthorhombic phase of CsPbI$_3$. 
(c) Rotated unit cell of CsPbI$_3$.
Color code: Cs (green), Pb (grey), and I (purple). 
(d) The thermodynamic stability diagram for CsPbI$_3$ and competing phases. The grey region denotes the thermodynamically stable chemical potential region for CsPbI$_3$.
}
\end{figure}

Figure~\ref{fig:phasediagram} (a-b) shows the unit cell of orthorhombic CsPbI$_3$, visualized using VESTA.\cite{Momma_2011_VESTAsoftware}
For the (001) surface, we build the slab model by constructing a 2 $\times$ 2 $\times$ 2.25 supercell from the unit cell, incorporating five CsI layers and four PbI$_2$ layers and a vacuum with a thickness of 20 \text{\AA} along the c-axis.
In the case of the (110) surface, the slab model was built by constructing a new unit cell defined by \(a'\) = \((-c)\), \(b'\) = \((b-a)\), and \(c'\) = \(a+b\).
The (110) slab is built by a 1 $\times$ 2 $\times$ 2.25 supercell based on the new unit cell shown in Fig.~\ref{fig:phasediagram} (c).
The CsI-terminated (110) surface contains five CsI layers and four PbI$_2$ layers.
For the slab model of the (100) surface, we reoriented the original crystallographic axes such that the new \(a'\) axis is aligned with \(-c\), the \(b'\) axis remains aligned with \(b\), and the \(c'\) axis is redefined as \(a\).
A 1$\times$2$\times$2 supercell is built for the (100) slab.
The Brillouin zone is sampled using a 1 $\times$ 1 $\times$ 1, 2 $\times$ 1 $\times$ 1, and 2 $\times$ 1 $\times$ 1 $\Gamma$-centered k-mesh for the (001) (110) and (100) surfaces.

\subsection{\label{sec_surface_energy} Surface energy and phase diagram}

To study the surface stability under various Cs, Pb, and I chemical potentials, we build surface structures with different stoichiometries including CsI-rich, PbI\(_2\)-rich, Cs-rich, Pb-rich, I-rich and stoichiometric surfaces. 
Identical structures are employed on both the top and bottom of the slab.
These surfaces are named using the total number of atoms in the supercell: \(\text{Cs}_\alpha \text{Pb}_\beta \text{I}_\gamma (\text{CsPbI}_3)_\delta\).
The surface energy (\(\Omega\)), which is also referred to as the grand potential\cite{sanna2010lithium_grand_potential}, is defined as:
\begin{align}
\Omega(&\Delta\mu_{\text{Cs}}, \Delta\mu_{\text{Pb}}, \Delta\mu_{\text{I}}) \notag \\
&= \{E_{\text{tot}}[\text{Cs}_\alpha \text{Pb}_\beta \text{I}_\gamma (\text{CsPbI}_3)_\delta] -\alpha \mu_{\text{Cs}}^{\text{bulk}} -\beta \mu_{\text{Pb}}^{\text{bulk}} \notag \\
&\quad - \frac{\gamma}{2} \mu_{I_2}^{\text{gas}} - \delta \mu_{\text{CsPbI}_3}^{\text{bulk}} -\alpha \Delta\mu_{\text{Cs}} -\beta \Delta\mu_{\text{Pb}} \notag \\
&\quad - \gamma \Delta\mu_{\text{I}} - \delta \Delta\mu_{\text{CsPbI}_3}\} / (2A).
\end{align}
\(E_{\text{tot}}[\text{Cs}_\alpha \text{Pb}_\beta \text{I}_\gamma (\text{CsPbI}_3)_\delta]\) is the total energy of the slab containing \(\alpha\) Cs atoms, \(\beta\) Pb atoms, \(\gamma\) I atoms, and \(\delta\) complexes of CsPbI\(_3\).
\({\mu}_i^{\text{bulk}}\) represents the energy of a Cs or Pb atom or an I\(_2\) molecule.
\(\Delta{\mu}_i\) represents the chemical potential of Cs, Pb, or I.
\(\Delta\mu_{\text{CsPbI}_3}\) is set to zero under the thermodynamic equilibrium condition of CsPbI$_3$.
\(\Delta\mu_{\text{Cs}}\), \(\Delta\mu_{\text{Pb}}\), and \(\Delta\mu_{\text{I}}\) are constrained by
\begin{align} 
\notag
\Delta\mu_{\text{Cs}} + \Delta\mu_{\text{Pb}} + 3\Delta\mu_{\text{I}} = \Delta H(\text{CsPbI}_3) = -5.77 \text{ eV} \\ \notag
\Delta\mu_{\text{Cs}} + \Delta\mu_{\text{I}} < \Delta H(\text{CsI}) = -3.30 \text{ eV} \\\notag
\Delta\mu_{\text{Cs}} + 3\Delta\mu_{\text{I}} < \Delta H (\text{CsI}_3) = -3.92 \text{ eV} \\\notag
\Delta\mu_{\text{Cs}} + 4\Delta\mu_{\text{I}} < \Delta H (\text{CsI}_4) = -4.16 \text{ eV} \\
\Delta\mu_{\text{Pb}} + 2\Delta\mu_{\text{I}} < \Delta H (\text{PbI}_2) = -2.41 \text{ eV}.
\end{align}
The shaded grey region in Figure~\ref{fig:phasediagram} (d) represents the thermodynamically stable region of CsPbI$_3$.

As \(\Delta\mu_{\text{Cs}}\) can be expressed as \(\Delta\mu_{\text{Cs}} = \Delta H(\text{CsPbI}_3) -\Delta\mu_{\text{Pb}} - 3 \Delta\mu_{\text{I}}\), \(\Omega\) can be defined as a function of \(\Delta\mu_{\text{Pb}}\), and \(\Delta\mu_{\text{I}}\) as
\begin{multline}
\Omega(\Delta \mu_{\text{Pb}}, \Delta \mu_{\text{I}}) = \{E_{\text{tot}}[\text{Cs}_\alpha \text{Pb}_\beta \text{I}_\gamma (\text{CsPbI}_3)_\delta] - \alpha \mu_{\text{Cs}}^{\text{bulk}} \\ - \beta \mu_{\text{Pb}}^{\text{bulk}} - \frac{\gamma}{2} \mu_{I_2}^{\text{gas}} - \delta \mu_{\text{CsPbI}_3}^{\text{bulk}} + (\alpha - \beta) \Delta \mu_{\text{Pb}} \\
+ (3\alpha - \gamma) \Delta \mu_{\text{I}} - \alpha \Delta H(\text{CsPbI}_3) \}/(2A).
\end{multline}

Based on this approach, we calculated the $E_{\text{tot}}[\text{Cs}_\alpha \text{Pb}_\beta \text{I}_\gamma (\text{CsPbI}_3)_\delta]$ for all surface structures.
The formation energies plotted as a function of $\Delta\mu_{\text{Pb}}$, and at a fixed value of $\Delta\mu_{\text{I}} = - 0.8$ eV are included in Fig. S1 in the Supporting Information.
We also developed a Python code to determine the surface that is most stable at a specific combination of \(\Delta\mu_{\text{Pb}}\) and \(\Delta\mu_{\text{I}}\) to generate the surface phase diagram. 
The code reads the formation energies calculated at \(\Delta\mu_{\text{Cs}}=\Delta\mu_{\text{Pb}}=\Delta\mu_{\text{I}}=0\) eV, $\alpha$, $\beta$, $\gamma$, $\delta$, and $A$ values from a .csv file, evaluates the most stable surfaces under the user-defined chemical potential region of \(\Delta\mu_{\text{Pb}}\) and \(\Delta\mu_{\text{I}}\), and generates a color contour plot for the phase diagram.

\section{\label{results}Results and Discussions}

In the following sections, we discuss the surface phase diagrams of (001), (110) (section \ref{sec001}), and (100) (section \ref{sec100}), and the structures of the most stable surface under the thermodynamic equilibrium condition of CsPbI$_3$.
In section \ref{combined}, we generate the phase diagram of all three surface terminations.
The electronic properties of the most stable surface under the thermodynamic equilibrium condition will be discussed.

\subsection{\label{sec001} Surface phase diagrams of (001) and (110) }

\begin{figure}
\includegraphics[width=90mm]{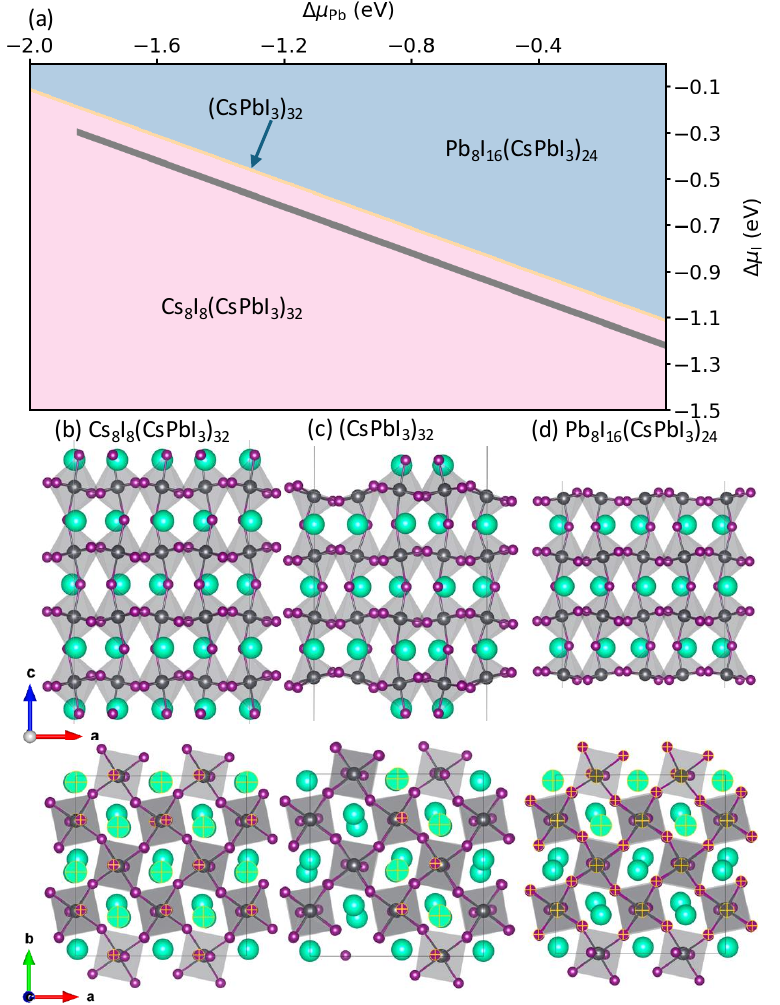}
\caption{\label{fig:001} (a) Surface phase diagram of (001). 
The region in grey denotes the thermodynamically stable region of bulk CsPbI$_3$. 
The relaxed surface structures of (b) Cs$_8$I$_8$(CsPbI$_3$)$_{32}$, (c) (CsPbI$_3$)$_{32}$, and (d) Pb$_8$I$_{16}$(CsPbI$_3$)$_{24}$.
}
\end{figure}

We created surfaces with 21 different stoichiometries for the (001) slabs.
For certain stoichiometries, multiple structures were generated to determine the most stable structure, resulting in a total of 25 structures. Additionally, 10 stoichiometries and 12 structures were calculated for the (110) surface.
These surface structures cover CsI-rich, PbI-rich, Pb-rich, Cs-rich, I-rich and stoichiometric surfaces.
For the (001) termination, we identified only CsI-rich, PbI-rich, and stoichiometric surfaces are stable in the surface phase diagram [Fig.~\ref{fig:001} (a)].

The surface with Cs$_8$I$_8$(CsPbI$_3$)$_{32}$ [Fig.~\ref{fig:001} (b)] is a CsI-terminated flat surface, which is stable under a wide range of Pb and I chemical potentials under CsI-rich conditions.
Moving to a less CsI-rich condition, the stoichiometric surface with (CsPbI$_3$)$_{32}$ [Fig.~\ref{fig:001} (c)] is observed in the phase diagram, which is stable under a narrow region of the Pb and I chemical potential.
Under PbI$_2$-rich conditions, the surface with Pb$_8$I$_{16}$(CsPbI$_3$)$_{24}$ [Fig.~\ref{fig:001} (d)] becomes stable, which is a PbI$_2$-terminated flat surface.
The surface energies in the unit of eV/cell and J/m$^2$ are summarized in Table~\ref{table:surface_energy}.
At the CsI-boarder of the thermodynamically stable region of bulk CsPbI$_3$ ($\Delta\mu_{\text{Cs}} + \Delta\mu_{\text{I}}=-3.3$ eV), the surface energy of Cs$_8$I$_8$(CsPbI$_3$)$_{32}$ is 0.062 J/m$^2$, which is consistent with the reported value (0.067 J/m$^2$).\cite{Yang_2022_cspbx3_surface}
At the PbI$_2$-boarder of the thermodynamically stable region of bulk CsPbI$_3$ ($\Delta\mu_{\text{Pb}} + 2\Delta\mu_{\text{I}}=-2.41$ eV), the surface energy of Pb$_8$I$_{16}$(CsPbI$_3$)$_{24}$ is 0.147 J/m$^2$, which is also close to the reported value (0.177 J/m$^2$).\cite{Yang_2022_cspbx3_surface}

As shown in the diagram, the thermodynamically stable region of bulk CsPbI$_3$ falls within the range of the Cs$_8$I$_8$(CsPbI$_3$)$_{32}$ surface, indicating that the CsI-terminated surface can be observed experimentally. 
Previous DFT calculations on the (001) surface of CsPbI$_3$ have also shown that the CsI-terminated surface is more stable than the PbI$_2$-terminated surface under the thermodynamically stable region of bulk CsPbI$_3$.\cite{Seidu_2021_cspbi3001surface_diffdefects}
Similar findings have also been reported for the (001) surface of orthorhombic CsPbBr$_3$ and CsPbCl$_3$.\cite{Yang_2022_cspbx3_surface}
This is also consistent with the experimental results reporting that the MA-halogen-terminated surfaces cover the majority of the (001) surface of MAPbBr$_3$ and MAPbI$_3$.\cite{Zhang_2022_stmreview_metalhalidepsc, QiSTMmapbbr, MAPbI3stm, zhongSTM}

The stoichiometric surface with (CsPbI$_3$)$_{32}$ can be constructed by creating vacancies on either CsI or PbI$_2$-terminated surfaces.
We find that the most stable stoichiometric surface is generated by creating four Cs vacancies and four I vacancies from both the top and bottom surfaces of the CsI-terminated slab, as shown in Fig.~\ref{fig:001} (c).
The surface energy is 2.14 eV/cell (0.108 J/m$^2$).
The stoichiometric surface can also be generated by removing four Pb and eight I atoms from both the top and bottom surface of the PbI$_2$-terminated slab [Fig. S2 (a)].
A similar structure was also reported by previous DFT studies.\cite{Haruyama_2014_mapbi3vacant_dftcal,haruyama2016mapbi3_surface}
This surface generated from the PbI$_2$-terminated surface is 0.14 eV/cell (0.007 J/m$^2$) higher in energy than the stoichiometric surface structure in Fig.~\ref{fig:001} (c).
Our results indicate that it is more likely to observe CsI-terminated surface and CsI vacancies on the CsI-terminated surface.
We note that MABr vacancies have been observed experimentally on the MAPbBr$_3$ (001) surfaces.\cite{liu2017mabrbr3_stm, stecker2019mabrbr3_stm}

\begin{figure}
\includegraphics[width=85mm]{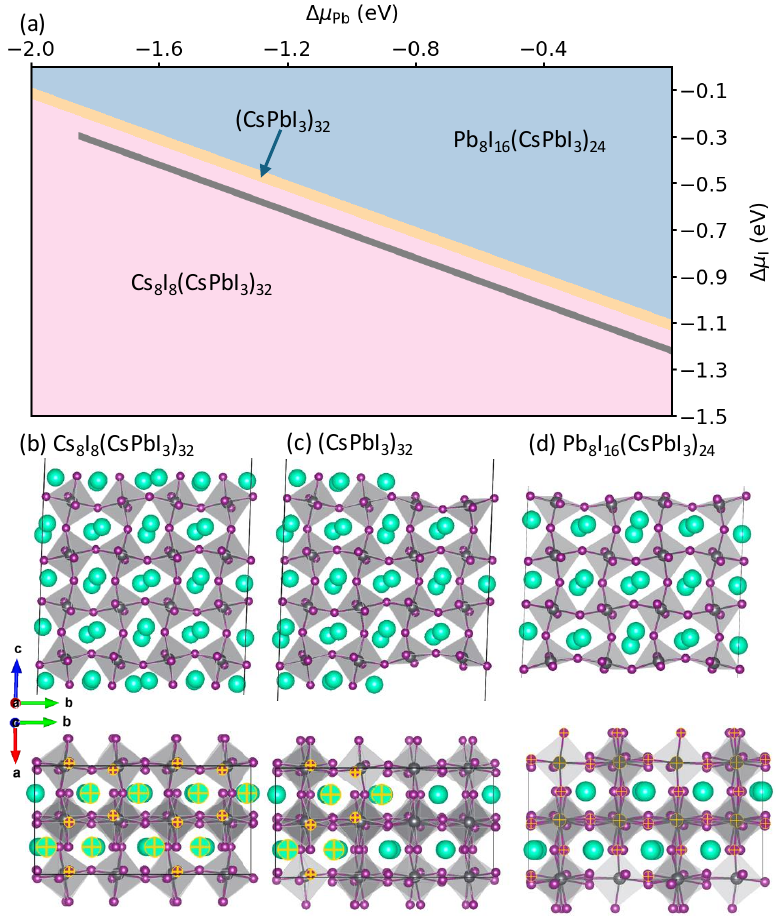}
\caption{\label{fig:110} (a)Phase diagram of the (110) surface. The thermodynamically stable region of bulk CsPbI$_3$ is marked in grey. 
Surface structures of (b) Cs$_8$I$_8$(CsPbI$_3$)$_{32}$, (c) (CsPbI$_3$)$_{32}$, and (d) Pb$_8$I$_{16}$(CsPbI$_3$)$_{24}$.
}
\end{figure}

\begin{table*}[htb]
\caption{\label{table:surface_energy}%
The surface energy ($\Omega$) in eV/cell and J/m$^2$, surface area ($A$), and the number of bonds broken on (001), (110), and (100) surfaces. 
$\Omega$ (J/m$^2$) is calculated by $\Omega$ (eV/cell) / $A$ and unit conversion (1 eV/${\mathrm{\AA}}^2=16.02$ J/m$^2$)
}

\begin{ruledtabular}
\begin{tabular}{cccccccc}
\textrm{ }&
\textrm{$\Omega$ (eV/cell)}&
\textrm{$A$ ($\mathrm{\AA}^2$)}&
\textrm{$\Omega$ (J/$\mathrm{m}^2$)}&
\textrm{Pb-I} (/cell)&
\textrm{Cs-I} (/cell)&
\\
(001)Cs$_8$I$_8$-(CsPbI$_3$)$_{32}$ & -11.97-4$\Delta\mu_\mathrm{Cs}$-4$\Delta\mu_\mathrm{I}$ & 317.187 & -0.604-0.202$\Delta\mu_\mathrm{Cs}$-0.202$\Delta\mu_\mathrm{I}$ & 8 (0.025)   & 24 (0.076)  \\
(001)Pb$_8$I$_{16}$-(CsPbI$_3$)$_{24}$ & -6.72-4$\Delta\mu_\mathrm{Pb}$-8$\Delta\mu_\mathrm{I}$ & 317.187 & -0.340-0.202$\Delta\mu_\mathrm{Pb}$-0.404$\Delta\mu_\mathrm{I}$ & 8 (0.025)  & 24 (0.076)  \\
(001)(CsPbI$_3$)$_{32}$ & 2.14 & 317.187 & 0.108 & 8 (0.025)  & 24 (0.076)  \\
(110)Cs$_8$I$_8$-(CsPbI$_3$)$_{32}$ & -12.00-4$\Delta\mu_\mathrm{Cs}$-4$\Delta\mu_\mathrm{I}$ & 318.565 & -0.604-0.201$\Delta\mu_\mathrm{Cs}$-0.201$\Delta\mu_\mathrm{I}$ & 8 (0.025)  & 20 (0.063)  \\
(110)Pb$_8$I$_{16}$-(CsPbI$_3$)$_{24}$ & -6.73-4$\Delta\mu_\mathrm{Pb}$-8$\Delta\mu_\mathrm{I}$ & 318.565 &-0.338-0.201$\Delta\mu_\mathrm{Pb}$-0.402$\Delta\mu_\mathrm{I}$ & 8 (0.025)  & 20 (0.063)   \\
(110)(CsPbI$_3$)$_{32}$ & 2.05 & 318.565 & 0.103  & 8 (0.025) & 20 (0.063)
\\
(100)(CsPbI$_3$)$_{24}$ & 1.37 & 230.662 & 0.095  & 8 (0.035) & 12 (0.052)
\\
(100)Cs$_8$I$_{8}$-(CsPbI$_3$)$_{20}$ & -11.71-4$\Delta\mu_\mathrm{Cs}$-4$\Delta\mu_\mathrm{I}$ & 230.662 &-0.812-0.278$\Delta\mu_\mathrm{Pb}$-0.278$\Delta\mu_\mathrm{I}$ & 8 (0.035)  & 20 (0.087)   \\
\end{tabular}
\end{ruledtabular}
\end{table*}

The bonding environment of the (110) surface is similar to that of the (001) surface: both surfaces consist of alternating CsI and PbI$_2$ layers along the $c$ or $c'$ direction. 
The surface phase diagram [Fig.~\ref{fig:110} (a)] and the stoichiometries of the stable surfaces [Fig.~\ref{fig:110} (b-d)] of (110) are therefore similar to (001).
The thermodynamically stable region of bulk CsPbI$_3$ also falls within the Cs$_8$I$_8$(CsPbI$_3$)$_{32}$ surface as shown in Fig.~\ref{fig:110} (a), indicating that it is more likely to observe the CsI-terminated (110) surface than the PbI$_2$-terminated (110) surface.

The surface energy of (110) is slightly lower than that of (001), as summarized in Table ~\ref{table:surface_energy}.
This can be understood by comparing the number of bonds that need to be broken to generate the surface structures.
Take the stoichiometric surface as an example.
The surface energy of (110) is 0.103 J/m$^2$, as compared to (001) with 0.108 J/m$^2$.
In bulk CsPbI$_3$, Cs is 8-fold coordinated and Pb is bonded to six I atoms forming corner-sharing Pb-I$_6$ octahedra.
To form the (CsPbI$_3$)$_{32}$ surface, 8 Pb-I bonds and 24 Cs-I bonds need to be broken on (001) while 8 Pb-I bonds and 20 Cs-I bonds need to be broken on (110).
Additionally, the (110) surface has a slightly larger surface area.
The density of bonds to be broken is lower on (110) than (001), explaining the slightly lower surface energy of (110).
It is important to use a supercell to investigate the surface structures of (001) and (110), as surface reconstructions may allow a lower surface energy.
Take the (110) surface as an example, the unreconstructed stoichiometric surface [Fig. S2 (b)] has a surface energy of (0.132 J/m$^2$), which is (0.029 J/m$^2$) higher than the 2$\times$1  reconstructed structure [Fig.~\ref{fig:110} (c)].

\subsection{\label{sec100} Surface phase diagram of (100) }

The (100) surface has a different bonding environment as compared to the (001) and (110) surfaces. 
The (100) slab is formed by stacking mixed Cs-Pb-I layers. 
It is then expected that the flat surfaces of (100) are terminated by mixed Cs, Pb, and I.
We also explored CsI-rich, PbI$_2$-rich, I-rich, Cs-rich, and Pb-rich surfaces on (100).
In total, we generated 9 surface structures with 7 different stoichiometries.

\begin{figure}
\includegraphics[width=85mm]{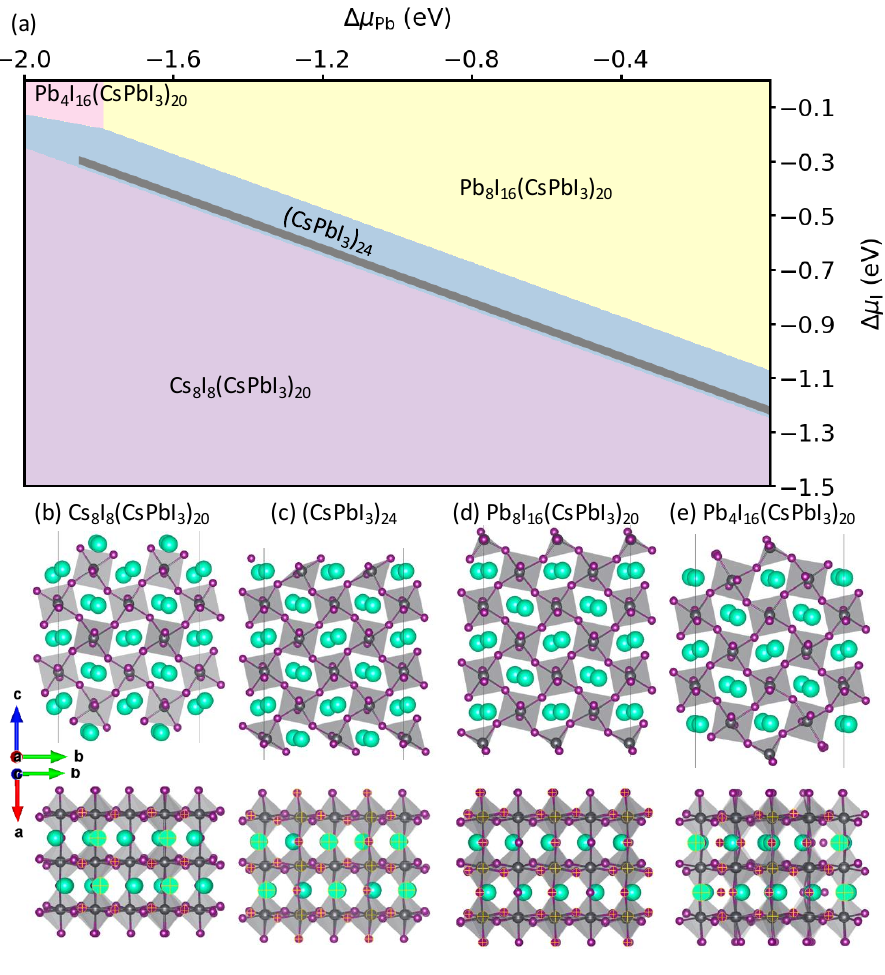}
\caption{\label{fig:(100)} (a) Surface phase diagram of (100). 
Surface structures of (b) Cs$_8$I$_8$(CsPbI$_3$)$_{20}$, (c) (CsPbI$_3$)$_{24}$, (d) Pb$_8$I$_{16}$(CsPbI$_3$)$_{20}$, and (e) Pb$_4$I$_{16}$(CsPbI$_3$)$_{20}$.
}
\end{figure}

Four surface terminations are stable in the surface phase diagram [Fig.~\ref{fig:(100)} (a)].
The surface with Cs$_8$I$_8$(CsPbI$_3$)$_{20}$ is stable under CsI-rich conditions while the surface with Pb$_8$I$_{16}$(CsPbI$_3$)$_{20}$ is stable under PbI$_2$-rich conditions, which are similar to (001) and (110) surfaces. 
What is unique about the (100) surface is that the stoichiometric surface on (100) is stable under a wide range of chemical potentials in the surface phase diagram, and the thermodynamically stable region of the bulk CsPbI$_3$ is found to be within the range of the stoichiometric surface, indicating that the flat stoichiometric (100) surface will be observed experimentally.
Under the Pb-poor and I-rich condition, we also found the surface with Pb$_4$I$_{16}$(CsPbI$_3$)$_{20}$ to be stable.
This surface is generated by creating two Pb vacancies on both the top and bottom surfaces of the Pb$_8$I$_{16}$(CsPbI$_3$)$_{20}$ slab.

We note that the surface energy of the stoichiometric surface of (100) is 0.095 J/m$^2$, which is the lowest among the three stoichiometric surfaces.
As there is growing interest in calculating surface defects and their roles in carrier lifetime and material stability\cite{long_2019_cspbi3001sur_csipbicompare_vacancys, Han_2022_passivationcspbi3_FtoI}, we propose the stoichiometric (100) surface for future studies of surface defect properties. 
It is a stoichiometric surface with surface energy independent of Pb, Cs, or I chemical potentials.
Additionally, it is a flat surface that can host Cs, Pb, and I vacancies, interstitials, and antisites.

\begin{figure}
\includegraphics[width=85mm]{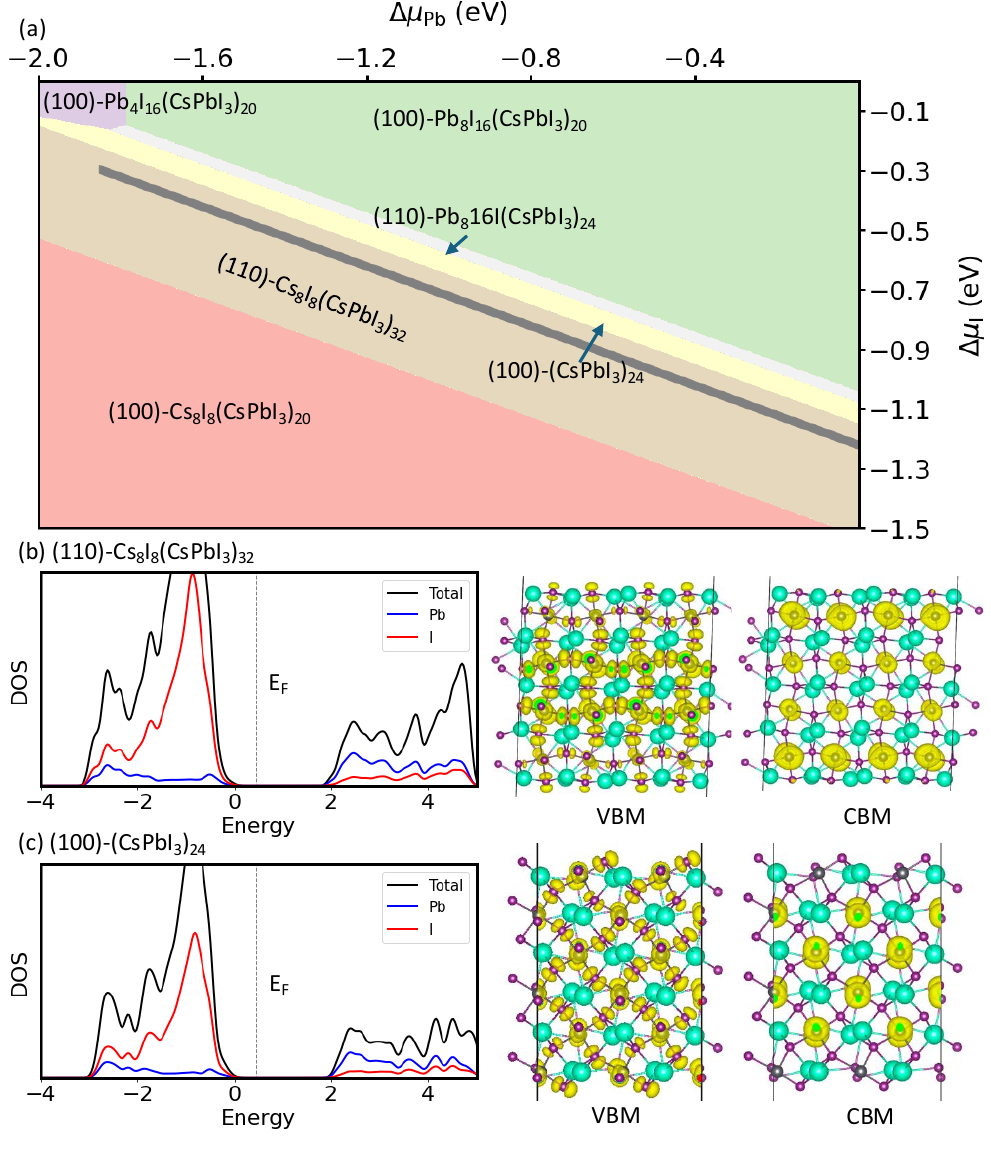}
\caption{\label{fig:allsurfaces} (a) Surface phase diagram of (001),(110) and (100). The thermodynamically stable region of bulk CsPbI$_3$ is in grey. 
The projected density of states and the partial charge density of the valence band maximum and conduction band minimum of (b) (110) Cs$_8$I$_8$(CsPbI$_3$)$_{32}$ and (c) (100)(CsPbI$_3$)$_{24}$. 
}
\end{figure}

\subsection{\label{combined} Surface phase diagram of (001), (110), and (100) and electronic structures}

We generate a surface phase diagram for all surface terminations as shown in Fig.~\ref{fig:allsurfaces} (a).
The thermodynamically stable region of the bulk CsPbI$_3$ falls within the Cs$_8$I$_8$(CsPbI$_3$)$_{32}$ surface on (110).
Moreover, it is noteworthy that the stoichiometric (CsPbI$_3$)$_{32}$ surface (100) is also close to the stable region of bulk CsPbI$_3$.
At the CsI border of the thermodynamically stable region of CsPbI$_3$, the surface energy of (110)-Cs$_8$I$_8$(CsPbI$_3$)$_{32}$ is 0.059 J/m$^2$ while the surface energy of (100)-(CsPbI$_3$)$_{24}$ is 0.095 J/m$^2$, suggesting that these two surfaces can co-exist.

The (001) surfaces do not show on the combined surface phase diagram mainly because, for the slab with the same stoichiometry, the (001) surface has a slightly higher surface energy (110). 
This phenomenon has been explained by counting the number of bonds to be broken to generate the corresponding surface structures (Table \ref{table:surface_energy}).
However, we note that the energy difference between (110) and (001) is small.

To better understand the electronic structures of the surfaces, we calculate the projected density of states of the Cs$_8$I$_8$(CsPbI$_3$)$_{32}$ surface on (110) and the (CsPbI$_3$)$_{32}$ surface on (100) in Fig.~\ref{fig:allsurfaces} (b-c).
The valence band is mainly contributed by the I-$p$ states while the conduction band is mainly contributed by the Pb-$p$ states.
Figure Fig.~\ref{fig:allsurfaces} (b-c) shows that there is no surface state above the valence band maximum (VBM) or below the conduction band minimum (CBM) of the Cs$_8$I$_8$(CsPbI$_3$)$_{32}$ surface on (110) and the stoichiometric surface of (100).
The DOS of these two surfaces shows that the surface electronic structures closely resemble the bulk property, indicating that these surfaces can contribute to the long carrier lifetime.

\section{\label{conclution}Conclusions}

In conclusion, we perform DFT calculations to study the stability of the (001), (110), and (100) surfaces of the orthorhombic CsPbI$_3$ considering various stoichiometries and reconstructions. 
We also develop Python code that can read the raw DFT data and plot the surface phase diagrams at user-defined chemical potentials, which illustrate the most stable surface under different combinations of Cs, Pb, and I chemical potentials.
Our discussions of surface properties are mainly focused on the thermodynamically stable region of bulk CsPbI$_3$.
The (001) and (110) surfaces have very similar bonding environments and similar surface phase diagrams.
We found the CsI-terminated flat surface is the most stable under the thermodynamically stable region of bulk CsPbI$_3$ for (001) and (110).
For the (100) surface, the stoichiometric surface [(CsPbI$_3$)$_{24}$] is the most stable under the thermodynamically stable region of bulk CsPbI$_3$.

We also generate a surface phase diagram across all (001), (110), and (100) surfaces.
The CsI-terminated flat surface [Cs$_8$I$_8$(CsPbI$_3$)$_{32}$] on (110) is the most stable under the thermodynamically stable region of the bulk CsPbI$_3$. 
Additionally, through the density of states and partial charge density analyses, we find no surface state on the Cs$_8$I$_8$(CsPbI$_3$)$_{32}$ surface on (110).
This suggests the surface electronic structure closely resembles the bulk properties, indicating good charge transport properties.

\section*{Data and code availability}
The datasets and code used to generate the surface phase diagrams are available from https://github.com/Mengen-W/pysurf.

\begin{acknowledgments}
The work was supported by the new faculty start-up and Transdisciplinary Areas of Excellence (TAE) Seed Grant funds from SUNY Binghamton.
Computing resources were provided by the Center for Functional Nanomaterials, which is a U.S. Department of Energy Office of Science User Facility, and the Scientific Data and Computing Center, a component of the Computational Science Initiative, at Brookhaven National Laboratory, which are supported by the U.S. Department of Energy, Office of Basic Energy Sciences, under Contract No. DE-SC0012704, and the Spiedie cluster at SUNY Binghamton. 
\end{acknowledgments}

\nocite{*}

\bibliography{Surface_phase_diagram_of_cspbi3}

\end{document}